\documentclass[12pt]{article}

\usepackage{graphicx}

\title{ The fundamental scale of QCD }

\author{Yu.A.Simonov \\
 NRC ``Kurchatov Institute'' -- ITEP,\\ B. Cheremushkinskaya 25, \\Moscow, 117218, Russia}

\newcommand{\beq}{\begin{eqnarray}}
 \newcommand{\eeq}{\end{eqnarray}}
\newcommand{\be}{\begin{equation}}
 \newcommand{\ee}{\end{equation}}

 \def\la{\mathrel{\mathpalette\fun <}}
\def\ga{\mathrel{\mathpalette\fun >}}
\def\fun#1#2{\lower3.6pt\vbox{\baselineskip0pt\lineskip.9pt
\ialign{$\mathsurround=0pt#1\hfil ##\hfil$\crcr#2\crcr\sim\crcr}}}

\newcommand{{\SD}}{\rm SD}

\newcommand{{\Mc}}{\mathcal{M}}

\newcommand{\llan}{\langle\langle}
\newcommand{\rran}{\rangle\rangle}
\newcommand{\lan}{\langle}
\newcommand{\ran}{\rangle}

\begin{document}
\maketitle
\begin{abstract}
There are several scales in the QCD as the theory of strong interaction: the vacuum gluonic condensate (as the divergence of the dilatation current), the nucleon mass as the basic mass scale in our universe, which is connected to the string tension $\sigma$, and the perturbative QCD scale $\Lambda_{\rm QCD}$, which defines the scale of
the renormalized perturbative expansions. In this paper we connect all these scales to the vacuum condensate, using the field correlator method, where the string  tension is expressed in terms of nonlocal field correlators, which are connected to the local condensates, while the perturbative $\Lambda_V$ is connected to $\sigma$ in the framework of the Background Perturbation Theory (BPT). We also demonstrate how the IR singularities and IR renormalons disappear in BPT making the resulting theory internally consistent.
\end{abstract}

\section{Introduction}
The basic scale of massless QCD can be related to the dilatation current anomaly \cite{0}.
Indeed, exploiting the divergence of the dilatation current
$\partial_\mu D_\mu= \Theta_{\mu\mu}$ in the massless QCD \cite{2}, one obtains a relation between $\Theta_{\mu\mu}$ and the gluonic condensate,
\be
\lan 0|\Theta_{\mu\mu}|0\ran= -\frac{\pi \beta(\alpha_s)}{2\alpha_s} \lan 0|G^2_{\mu\nu}|0\ran,
\label{eq.1}
\ee
and finally  the vacuum energy density $\epsilon_V$ \cite{2},

\be
\epsilon_V= -\frac{\pi \beta(\alpha_s)}{8\alpha_s} \lan 0|G^2_{\mu\nu}|0\ran.
\label{eq.2}
\ee
The analysis of the Operator Product Expansion allows to connect the effective QCD mass parameters to the
gluon condensate $G_2= \frac{\alpha_s}{\pi} \lan 0|G^2_{\mu\nu}|0\ran$ \cite{2,3}. At this point one can ask how
 $ G_2$ can be connected to another - ``phenomenological" basic scale - the nucleon mass $M_N$? Indeed, $M_N$ with the value around 1 GeV defines
the most masses  in the universe and it is not clear  how $M_N$ is connected to $\epsilon_V$. Here one might remember the basic property of QCD -- the confinement
\cite{3,4,5,6} and its fundamental scale - the string tension $\sigma$, well defined both experimentally and in theory (see a detailed discussion in \cite{6}).
Now it is understood that the nucleon mass is defined by $\sigma$ --this topic is well studied and the baryon masses, including the nucleon mass, are
calculated with a good accuracy in the QCD framework (see e.g. \cite{6*}). As a result one meets with the basic
question: how $\sigma$ is connected to the gluonic condensate $G_2$ ?  This topic will be the main point of the present paper with the aim to find the explicit
numerical relation between $\sigma$ and $G_2$. And moreover, one puts another basic problem: the perturbative QCD (pQCD) has its own basic scale -- $\Lambda_{QCD}$,
which defines the renormalized amplitudes \cite{2}.  Is it another scale or it can be connected to the previous scales, $\sigma,~G_2$?
It is the purpose of the paper to discuss these questions and give  our answers.

In this discussion our main point is theory of confinement in QCD, based on the field correlators (FC) \cite{3,4,5,6},
which is supported by lattice and experimental data and is the basis of all strong interactions in QCD. Being
internally consistent it gives an explicit mechanism of confining force and the string tension $\sigma$ via the field correlators and explains hadron masses, including the proton mass, via $\sigma$. However,  the connection between $\sigma$ (and hadron masses on one side) and the gluonic condensate $G_2$ was not found  before (outside of the sum rule approach \cite{2,3}), and below this will be given for the first time. The second important topic is the scale of pQCD, associated with
$\Lambda_{QCD}$, which usually is treated as an independent scale. Below we study the possible connection of this
scale with the ``strong scales" -- $\sigma,~ G_2$. In doing so we are exploiting the Background Perturbation Theory (BPT)
with the theory of confinement as the proper background and find the evidence for the agreement of the BPT scale with
the confinement scale. Finally we show that confinement plays the basic role in eliminating the IR singularities and
the IR renormalons within the BPT, which can make the resulting theory fully internally consistent.

The paper is organized as follows. In next section we show how the  string tension $\sigma$ is calculated via the FC and the gluelumps.
In the course of derivation we obtain that $\sigma$ is directly connected to the gluonic condensate and moreover, how the vector coupling $\alpha_V(1.0$~GeV$^{-1})$ in the coordinate space
can be estimated from the resulting equations. In the section 3 we discuss the role of $\sigma$ in the BPT, which allows to extend pQCD to
smaller momenta (to larger distances) and avoid the Landau virtual pole problem. In section 4 the BPT based resolution of the IR renormalon problems and in general the IR divergence problems are shortly discussed.

\section{The string tension via the field correlators and the gluelump masses}

The basic property of confinement in QCD can be derived from the Wilson loop integral, containing the gluon fields
$A_\mu(z), F_{\mu,\nu}(z)$,

\be
W(C) = \frac{1}{N_c} \lan tr P\exp(ig \int_C dz_\mu A_\mu (z) )\ran = \frac{1}{N_c} \lan tr P\exp(ig \int_{S_{\min}} d\sigma_{\mu\nu}  F_{\mu\nu} )\ran.
\label{3}
\ee
One can apply to (\ref{3}) the operator cluster expansion \cite{6}, which allows to expand in the exponent the connected terms, thus producing connected correlators $\llan\rran$.
As a result the vacuum averaging over fields $F_{\mu\nu}$ yields

$$ W(C) = \frac{1}{N_c} tr \exp \left[ - \frac{g^2}{2} \int d\sigma_{\mu\nu} d\sigma_{\lambda\rho}\llan F_{\mu\nu} F_{\lambda\rho}\rran + \right.$$
\be \left.+ \frac{g^4}{4!} \int d\sigma (1) d\sigma(2) d\sigma(3) d\sigma(4)  \llan \hat F(1)\hat F(2) \hat F(3)\hat F(4)\rran +O(g^6)\right].
\label{4}
\ee
At this point one can introduce the quadratic field correlators $D^{(2)}$ \cite{4,5,6}, which establish the main properties of confinement, and in our case
we are interested in the colorelectric confinement, yielding  the mass to all hadrons,

$$  g^2 D^{(2)}_{i4k4} (x-y) \equiv \frac{g^2}{N_c} \lan tr_f (F_{i4} (x) \Phi(x,y) F_{k4} (y)
\Phi(y,x)\ran = (\delta_{ik} ) D^E(x-y)+$$ \be + \frac12 \left(
\frac{\partial}{\partial x_i} [h_k + {\rm ~perm}]\right) D_1^E (x-y), ~~
h_\lambda = x_\lambda -y_\lambda, ~~(x-y)^2 = \sum^4_{\lambda=1} (x_\lambda-y_\lambda)^2.
\label{5}
\ee
Here the parallel transporter $\Phi(x,y)= \int^y_x dz_\mu A_\mu$, connecting the points $x,y$ and ensuring
gauge invariance of the whole expression, enters. Insertion (\ref{5}) into (\ref{4}) gives the area law of the Wilson  loop,

\be
W(C) =\exp (-\sigma RT_4), ~~\sigma =\frac12 \int d^2z D^E(z).
\label{6}
\ee

We now turn to calculation of the static potentials, generated by $D^E,~D_1^E$ \cite{4,5,6,7,8}. We start with the Wilson loop,
$W(C) = \exp  \left(-\frac{g^2}{2} \int d\sigma \int d\sigma \lan F  F\ran\right)= \exp  \left (-\int (V_D^R) + V_1(R))dt_4\right)$
and obtain for the fundamental charges
\be
V_D (R) = 2 \int^R_0 (R-w_1) dw_1 \int^\infty_0 dw_4 D^E  \left(\sqrt{w^2_1+ w^2_4}\right)= V_{\rm conf} (R) + V_D^{\rm sat} (R).
\label{7}
\ee
As a result, one defines the string tension $\sigma$ as the integral of $D^E$ in the $(14)$ plane
\be
\sigma= 2 \int^\infty_0 dw_1 \int^\infty_0 dw_4 D^E \left(\sqrt{w^2_1+ w^2_4}\right)= \pi \int z dz D^E(z).
\label{8}
\ee
Till now we have considered only quadratic in $F_{\mu\nu}$ correlators, neglecting quartic and higher order terms. As it is shown in \cite{5,6}, the contribution of the correlators, containing product of $2n$ fields, is damped as $a^n$, where $a^{-1} = 16-20$, and therefore the accuracy of this formalism is around few percent.
At this point one must find what is the basic object which defines confinement in this leading approximation.
As shown in \cite{7,8,9}, one can obtain this basic object, keeping in (\ref{5}) in both fields $F$ the quadratic in $A_\mu$ terms,
which produce so-called gluelump Green's functions,
\be
 D^E(x-y) = \frac{g^4 (N_c^2-1)}{2} G^{(2g)} (x,y)= g^4 N_c C_2 (f)  G^{(2g)}(x,y).
\label{9}
\ee
The spectrum of the gluelumps $G^{(2g)}$ was calculated both analytically \cite{10} and on the lattice \cite{11}, and in \cite{12,13} the asymptotics was found as

\be
G^{(2g)} (x)  \approx 0.108 ~\sigma^2 e^{-M_0^{(2g )}|x|},~~ x\ga (M_0^{ (2g )} )^{-1},
\label{10}
\ee
where the mass of the two-gluon gluelump is  $M_0^{(2g)} \approx 2.5$ GeV, \cite{10,11}, $ (M_0^{ (2g )} )^{-1} \approx 0.1$ fm. The  mixing of two-gluon and
one-gluon (with the mass $M^{(1g)}$) gluelumps and the account of the color Coulomb interaction give rise to  the lowering of $M^{(2g)}$. Then we denote
$M_0^{(2g)}\equiv M_0$ and define it, using results from \cite{10,11} with account of color Coulomb force, and  express it via $\sigma$ \cite{7,8,9},

\be
M_0= 2.4~{\rm GeV} = 5.65 \sqrt{\sigma}.
\label{11}
\ee
Using (\ref{11}), one can write the asymptotic behavior of $D^E(z)= D_{\rm as}(z)$, $z >> \frac{1}{M_0}= 0.08$~fm. It is valid for $z > \lambda_0=1.0$~GeV$^{-1}$ with accuracy better than $10\%$. We shall use this point $\lambda_0$ as a meeting point between the asymptotic form $D_{\rm as}(z)$ and the small
$z$ form of $D^E(z)$, below denoted as $D(z)$,

\be
D(z)= D_{\rm as}(z)= 8 \pi^2 (N_c^2-1) 0.108 \alpha_V^2 \sigma^2 \exp(-M_0 |z|).
\label{12}
\ee

\begin{figure}[!htb]
\begin{center}
 \includegraphics[width=45mm,keepaspectratio=true]{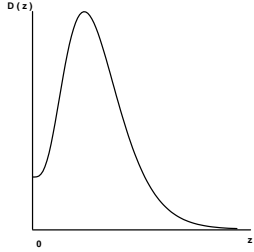}
 \caption{The colorelectric correlator$ D^E(z)$, defined from the gluelump Green's function at larger $z$ and with account of the radiative corrections at lower $z$,
both regions are connected at the peak $z=1.0$~GeV$^{(-1)}$.}
\end{center}
\end{figure}

Notice, that the correlator (\ref{12}), which defines the hadron properties, contains the vector coupling $\alpha_V(r)$ in the coordinate space
(see Appendix{A1.}). As one can see in Fig 1, the integral over $dz$ in (\ref{8}), defining $\sigma$, has the maximum
near $z=\lambda_0  \approx 0.2$~fm, which defines the parameter $\alpha_V(\lambda_0)$  to be discussed in what follows.

We now turn to the small distance contribution to $D(z)$, which was calculated in \cite{7,8,9,12,13} (see
 Fig. 2,3 and discussion in \cite{9}),
\be
D(z)= -4 N_c \alpha_V^2(z) G_2 + N_c^2 \frac{\alpha_V^2(z)}{2\pi^2} D(z_{\rm max}) \ln^2\left(\frac{z_{\rm max} \sqrt{e}}{z}\right),
\label{13}
\ee
where $z_{\rm max}$, defined by the $O(\alpha)$ correction due to one-gluon exchange between two gluons in the gluelump (the diagram in Fig 3 in \cite{9}), is of the order or less
$\sigma_{\rm adj}^{-1/2}= 0.31$~fm and we take it at the meeting point $z= \lambda_0 =1.0^{-1}=0.20$~fm and $\alpha_V(z\rightarrow 0)=\frac{2\pi}{\beta_0 \ln(\Lambda_V z)^{-1}}$. One can see that at small $z$ the first term is going to zero,
while second term on the r.h.s. dominates and tends to the constant $D(0)$ at $z= 0$ \cite{7,8,9}, equal to
\be
D(0)= \frac{N_c^2}{2\pi^2} D(\lambda_0) \left(\frac{2\pi}{\beta_0}\right)^2.
\label{14}
\ee
From (\ref{14}) one has the relation $D(0)= 0.15 D(\lambda_0)$, while $D(0)$ is connected to the gluonic condensate,
$D(0)= \frac{\pi^2}{18} G_2,~G_2= \frac{\alpha_s}{\pi} \lan 0|F_{\mu\nu}^a F_{\mu\nu}^a|0\ran$.

We can now compute the total string tension, by integrating $D(z)$ in the whole interval of $z$ as in (\ref{8}),
including internal and external regions of integration,
\be
\sigma= \Sigma_{in} + \Sigma_{ext}, ~~\Sigma_{in} = \int^{\lambda_0}_0, ~\Sigma_{ext}= \int^\infty_{\lambda_0},
\label{15}
\ee
where
\be
\Sigma_{in}= \pi \lambda_0^2 \frac{1}{3} D(\lambda_0) = \eta \lambda_0^2 \frac{1}{3} \exp(-M_0 \lambda_0) ,~~\Sigma_{ext} = \eta  \frac{1}{M_0^2} \exp(-M_0 \lambda_0)(1 + M_0\lambda_0),
\label{16}
\ee
Moreover $D(\lambda_0)=D_{\rm as}(\lambda_0)$, the latter
is defined in (\ref{12}), while $\eta = 214.3 \sigma^2 \alpha_V^2$.

At this point one can choose two different strategies: step 1 -we calculate the string tension $\sigma$ from both, internal and external regions, and putting it equal to the standard value $0.18 GeV^2$,
define the resulting value of $\alpha_V(\lambda_0)$. Step 2 : we define $D(\lambda_0)$ from the continuity condition with external value of
$D_{\rm as}(\lambda_0)$ and then find the resulting value of $G_2$.

{\bf Step 1.}

In this case from (\ref{16}) and the equation $\sigma= \Sigma_{in} + \Sigma_{ext}$ one obtains
\be
\sigma= \eta (\lambda_0^2 1/3 + \frac{1}{M_0^2}), ~~ \alpha_V^2(\lambda_0)= \frac{1}{A}, ~~
 A= 214 \exp(-M_0 \lambda_0) \left(1/3 \sigma \lambda_0^2 + \frac{\sigma (1+ \lambda_0 M_0)}{M_0^2}\right).
\label{17}
\ee
As a result, one has the optimal value of $M_0=2.4$~ GeV and within 10\% accuracy  the coupling  $\alpha_V(1.0$~GeV$^{-1})= 0.567$.
These numbers will be discussed in the next section, but now we turn to the Step 2.

{\bf Step 2.}

One  takes into account the continuity condition: $D_{\rm in}(\lambda_0)= D_{\rm as}(\lambda_0)$, which gives
\be
D(0)= 0.15 D(\lambda_0)= 0.15 64 \pi^2 \alpha_V^2 \sigma^2 \exp(-M_0 \lambda_0).
\label{18}
\ee
Then we determine the  value of the gluonic condensate $G_2$, defined by the relation $D(0)= \frac{\pi^2}{18} G_2$ , namely,

\be
G_2= 1.69 \sigma^2 \alpha_V ^2= 0.054 \alpha_V^2 {\rm GeV}^4.
\label{19}
\ee
This relation allows to find $G_2$ as the basic scale, corresponding to  measured string tension $\sigma$ and
the value of $\alpha_V(\lambda_0)$. Taking the optimal value of $M_0=2.4$ GeV, one obtains $G_2= 0.017$~GeV$^4$. This value is in the same ballpark as  the
standard value $G_2= 0.012$~GeV$^4$, exploited in the numerous sum rule analysis, see e.g. \cite{1}.

\section{The perturbative QCD scale vs the confinement scale}

Till now we have simplified our analysis, considering $\alpha_V$ as a constant, which does not change much for
$r=\lambda$ around $\lambda_0= 1.0$~GeV$^{-1}$. Also our  analysis in Step 1, where both internal and external (asymptotic) regions of $z$ were considered,
has given in (\ref{17}) reasonable value of the  coupling $\alpha_V(\lambda_0) = 0.567$.  As we shall see below
this value is close to  $\alpha_V(2-loop,r=0.20$~fm), defined via the 2-loop coupling in the momentum space $\alpha_V(Q^2)$ in the framework of BPT
(see Appendix A1.), in which the QCD constants $\Lambda_V(n_f)~(n_f=3.4.5)$ are used in different $Q^2$-regions and the IR regulator
$M_B=1.15$~GeV is used. In particular, $\Lambda_V(n_f=3)=500(15)$~MeV corresponds to $\Lambda_{\overline{MS}}(n_f=3)= 338(10)$~MeV, well established
in experiment \cite{16} and in lattice QCD \cite{17}. In this way we can estimate the pQCD scale  via
$\sigma$ and $G_2$, discussed in the previous section.

In this section we consider the couplings $\alpha_V(Q^2)$ in the momentum space and $\alpha_V(r)$ in the coordinate space in more detail, aiming at the connection
between standard perturbative and nonperturbative (BPT) aspects of QCD. Our final goal is to understand the meaning of the value of $\alpha_V(r= 0.2 $ Fm $ )=0.567$, calculated in the previous section.
We start with the value of $\alpha_V(Q)$ with $Q\cong 1.0$~GeV and ask ourselves whether one can use the standard perturbative representation of $\alpha_V(Q)$ in this region. Here
we can use the same arguments as in \cite{13,14,15,16,17,18}. To simplify matter we consider the limit $N_c\rightarrow \infty$, which implies that all gluon (adjoint) lines become
double fundamental lines, and therefore, considering arbitrary quark-antiquark interaction diagram, one will see closed fundamental lines in the gluon self-energy diagrams -- it
means that all closed contours are covered with fundamental film with the fundamental string junction $\sigma_f= 0.18$ GeV$^2$. Therefore the self-energy gluon part to the lowest
order in $\alpha_V$ can be written as the spectral sum over all $f-\bar f$ (here $f$ means fundamental) radially excited bound states with masses $M_n^2$, linearly
growing with $n$ : $M_n^2= M_B^2 + n m^2$  \cite{18*,18**}, where
$M_B^2= 2\pi \sigma$ and $m^2$ can also be expressed via $\sigma$, but disappears in the renormalization process.
\be
 \Pi^{(0)}(Q^2)= \sum_n^{\infty_0 }\frac{C_n}{Q^2 + M_B^2 + n m^2} = -\frac{N_c}{12\pi^2} \psi\left(\frac{Q^2 + M_B^2}.
 {m^2}\right).
 \label{20}
 \ee
 In (\ref{20}) we have omitted the divergent constant and  using the asymptotics of the $\psi-$function,
 \be
 \psi(z)= lnz -\frac{1}{2z} - O\left(\frac{1}{z^2}\right),
 \label{21a}
 \ee
 arrive at the asymptotic estimate of the gluon loop on the gluon line inside the quark-antiquark interaction diagram \cite{18*,18**},
 \be
 \Pi^{(0)}(Q^2)= - \frac{N_c}{12\pi^2} ln\left(\frac{Q^2+M_B^2}{m^2}\right) + O(1/Q^2).
 \label{22a}
 \ee
Extending this analysis to higher orders in $\alpha_V(Q)$  \cite{14,15,18}, one arrives at the expression for $\alpha_V(Q)$,
which in pQCD  without the confining background, is looking as
\be
\alpha_V(Q)= \frac{4\pi}{b_0 t} \left(1 - \frac{b_1 ln t}{b_0^2 t}\right) + ...,~~ t= ln\left(\frac{Q^2}{\Lambda_V^2}\right),
\label{21}
\ee
(see \cite{18***} for a fourth order $\alpha_V$ calculation )
and e.g. for $\Lambda_V(n_f=3)=500$~MeV the 2-loop coupling, $\alpha_V(Q=1.0~{\rm GeV}^{-1})=0.820$ is large. However, with account of confined gluon loops
(which become at large $N_c$ the fundamental loops) the expression (\ref{21}) keeps its form, but now
$t$ is replaced by $t_B$, where $t_B= ln\left(\frac{Q^2 + M_B^2}{\Lambda_V^2}\right)$. More generally this form is applicable  with any number of terms in the situation, when the pQCD
interaction in the system is considered on the confinement background, not connected with perturbative interaction. This question can be formulated as the perturbation theory in
the nonperturbative  vacuum, or more explicitly the perturbative series in the vacuum with the confining background, called BPT.

This topic has a long history and starts with the papers \cite{19,20,21} and later was developed explicitly for the confining background \cite{18*,18**,22,23,24}.
As it was shown there and developed further in \cite{14,15,18}, the basic equation (\ref{20}) should be replaced by the new equation, which has the same form but
the factor $t$ is equal to $t_B= ln\left(\frac{Q^2 + M_B^2}{\Lambda_V^2}\right)$ with the additional constant $M_B^2$ . The replacement of this kind (with $M_B$ equal
some arbitrary constant) was suggested earlier without rigorous derivation in \cite{26}, but in \cite{15}
this form was compared  with experimental data \cite{16} and the lattice predictions \cite{17}.
The basic point here is the equivalence of the gluon-loop contribution in the standard Gell-Mann-Low expression
and the $q-\bar q$ Green's function, expressed as the infinite sum over bound states. Interestingly, both have the
same logarithmic asymptotics at large $Q$, but the Green's function contains additional $M_B^2$ in $ln(Q^2 +M_B^2)$,
which allows to extend $\alpha_V$ to the momentum $Q^2\cong M_B^2$, where it agrees well with experimental and lattice results \cite{15}. And here we can preview that the pQCD can become a
self-consistent theory  only in the confining vacuum, so all parameters of QCD are to be defined with account of confinement.
The basic point here is that in the confining vacuum the gluon loop can be calculated both for positive and negative
$Q^2$ and  one obtains the logarithmic asymptotics at large positive $Q^2$, but one can analytically continue the results from positive to negative values of $Q^2$ and define the
BPT gluon loop in the whole $Q^2$ plane. One can ask whether the  old argument due to F.~Dyson     \cite{26*} of the essential singularity in the $g^2$ plane at $g^2=0$ holds also in BPT.
Indeed, the basic confinement quantities $\sigma, D(z)$ are proportional to $g^4$ and formally do not hold the Dyson arguments, but the difference seems to be more deep and calls for additional studies.

We now come to the final point of our investigation -- the comparison  $\Lambda_V(n_f=3)$, known both from lattice \cite{17} and experiment \cite{16},
with that, obtained from $\alpha_V(r=0.20$~fm$)= 0.567$, defined above in (\ref{17}). This procedure is described in detail in the Appendix A1. and consists of several steps:
1. the connection between  $\alpha_V(r)$ and $\alpha_V(Q)$, which requires the separation of the whole $Q$ interval in 3 regions of different $Q$ values and definition of
the overall (``the compound") $\alpha_V^c(Q)$. The resulting $\alpha_V^c(Q)$ is shown in Fig. 2 for two values of $M_B= 1.15, 1.0$ GeV. The calculations show that
$\alpha_V(r,n_f=3)$ almost coincides with the compound $\alpha_V^c(r)$, as it is seen in Fig. 3, which allows to define
$\Lambda_V(n_f=3)= 492$ MeV. Note that the exploited values of $M_B=1.0,~ 1.15$ GeV are close to the predicted above theoretical estimate $M_B^2= 2 \pi \sigma= 1.13 $ GeV.

This result of the BPT analysis allows to compare calculated $\alpha_V(0.20\rm fm)$ with our basic result - $\alpha_V(\lambda_0=0.20\rm fm)=0.567$ in Eq.~(\ref{17}).
Then using $\Lambda_V(n_f=3)=492$~MeV and the connection, $\Lambda_V(n_f=3)= 1.4753 \Lambda_{\overline{MS}}(n_f=3)$ (see Appendix A1.) one finally has $\Lambda_{\overline{MS}}(n_f=3)= 333$ MeV,
which agrees well with the lattice number $332(19)$ MeV \cite{17} and that from the analysis of experiment data in \cite{16}.

\begin{figure}[!htb]
\vspace{3ex}
\begin{center}
 \includegraphics[width=60mm]{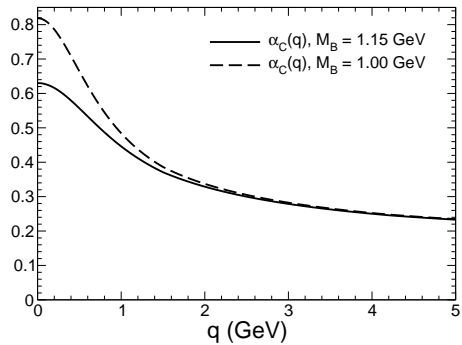}
\caption{The compound two-loop vector coupling $\alpha_V(Q)$ for $ M_B=1.15$ GeV
(solid line) and $M_B=1.0$ GeV (dashed line). In both cases $\Lambda_V(n_f=5)=310$ MeV, $\Lambda_V(n_f=4)=421$~MeV,
$\Lambda_V(n_f=3)=497$ MeV from \cite{15}.
\label{fig.2}}
\end{center}
\end{figure}

\begin{figure}[!htb]
\vspace{1ex}
\begin{center}
 \includegraphics[width=60mm]{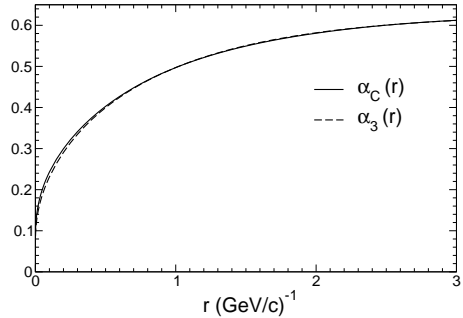}
\caption{ The compound coupling $\alpha_V(r)$ and the coupling $\alpha_V(r,n_f=3)$
for $M_B= 1.15$ GeV as functions of $r$ (other parameters as in Fig. 2), from \cite{15}.
\label{fig.3}}
\end{center}
\end{figure}

From Fig. 2 one can see that $\alpha_V(r= 0.2$~fm) is close to the value $\alpha_V(\lambda_0)\cong 0.56$ found in (\ref{17}) and in this way we can associate this value
with $\Lambda_V(n_f=3)= 497$ MeV. From here one finds the agreement with the standard definitions of $\Lambda_{\overline{MS}}= 333$ MeV and this means that the standard perturbative scale of QCD is correctly estimated from the confinement scale with the use of the BPT.

\section{Discussion}

In our analysis we have shown that the confinement theory, based on the FC method, allows to establish  the only QCD scale, $\sigma$ or $G_2$, for all strong interactions. In addition, using the  BPT,
one can extend this nonperturbative scale into the whole QCD region, suggesting a unique scale for all processes in QCD, with exception of the quark masses, anomalies etc.

But this is not a whole story -- indeed, the pQCD without confining background is a subject of many inconsistencies in the treatment of the sum of the perturbative series. Among them the most important  are the IR renormalon problem \cite{2,28,29} and the IR divergencies in the perturbation theory \cite{30,31,32,33,34,35}. The internal inconsistency of the standard pQCD, which follows e.g. from the existence of infrared renormalon singularities \cite{2,28}, stems possibly from the the same source as the Landau ghost pole problem, discussed in section 3, and  can be cured by the BPT similarly to the approach from \cite{29}. Indeed, the infinite set of gluon loops, interconnected by gluon lines (the standard IR renormalon diagrams), can be summed up both in standard PT and in the BPT as follows,
\be
\Delta\Pi(Q^2)= {\rm const} \sum_n\left(\frac{b_0\alpha_s(Q^2)}{4\pi}\right)^n\int^{Q^2}_0 \frac{k^2 d k^2}{Q^4}ln^n\left(\frac{Q^2+ m^2}{k^2+m^2}\right).
\label{22}
\ee
Here the BPT representation of the gluon loop with $m^2$ is used as in the (\ref{22a}), while in  the standard PT one should put $m^2=0$.
As a result of integration, one obtains, as in \cite{29},
\be
\Delta\Pi(Q^2)={\rm  const} \sum_{n>>!} \left(\frac{\alpha_s(Q^2)b_0}{8\pi}\right)^n q_n.
\label{23}
\ee
Here $q_n= \Gamma(n)$ for $n< 2 n_0$ and $q_n= \frac{(2n_0)^n}{n}$ for $n> 2n_0$ with  $n_0= ln(\frac{Q^2+m^2}{m^2})$.
One can see a drastic change in the situation when $m=0$ and one obtains in (\ref{20}) the Borel non-summable $n!$
series, while for nonzero $m$ one has a summable geometric series. In this way the BPT corrects the dangerous IR
renormalons. This topic can be extended with the hope to obtain in all cases  the summable perturbative series with the help of the BPT.

Another important property of the BPT is elimination of the IR singularities, appearing in the calculation of the loop
diagrams with zero mass propagators \cite{30,31,32} within the standard pQCD. Different ways of subtraction of diverging terms are introduced in the literature to obtain  finite terms subject to subsequent summation. The nonperturbative approach to parton and rescattering processes, suggested in \cite{33,34,35}, treats all quark and gluon loop integrals as integrals of the Wilson loops covered with fundamental or adjoint confining film. In the limit of large $N_c$ all these loops are disconnected and can be computed separately. As an example we give below the integral of the confined triangle diagram, computed in Appendix 2 of \cite{35},
 $$
 G(p_1,p_2,p_3)= tr \prod^3_1 \int d^4 q_i \Gamma_i (m_i - i q_i^a\gamma_a) \int^\infty_0 d s_i \exp(-s_i(m_i^2 +q_i^2)) I_3,$$
 \be I_3= (2\pi)^4 \delta(\sum p_i) 16 \pi^2 \exp(- \frac{2 \sqrt{b_1^2 b_2^2 -(b_1 b_2)^2}}{\sigma}.
 \label{24}
 \ee
 The most interesting property of $I_3$ is its limit for $\sigma= 0$,
 \be
 I_3 ~ \delta^{(4)}(b_1) \delta^{(4)}(b_2), b_1= -p_1 -p_3 -q_1 + q_2, b_2= p_3 -q_2 + q_3.
 \label{25}
 \ee
 As a result instead of three integrations over $dq_i$ in (\ref{24}), which make  the confined triangle IR nonsingular, in the $\sigma=0$ limit due to (\ref{25}) one has the only $d^4 q$
 integral and the standard IR singularity.  This mechanism is easily extended to higher loops and effectively makes all loop integrals (at least in the large $N_c$ limit) IR convergent.

 \section{Conclusions}

 We have shown that the basic element of confinement in the Field Correlator Method  -- the gluelump
 Green's function -- allows to compute the string tension $\sigma$ and at the same time to connect $\sigma$ to
 the basic element - the gluonic condensate $G_2$ and moreover, to estimate $\alpha_V$ at the scale of 0.2~fm.
 Using the latter value, one can estimate the intrinsic perturbative scale $\Lambda_{\overline{MS}}$ and establish an agreement with its value, well known from lattice and
 experimental data. In this way it becomes possible to define
 with some accuracy a unique scale in the QCD with confinement (e.g. $G_2$ or $\sigma$), which (together with the quark masses, anomalies etc) are responsible for
 all  scales in physical perturbative and nonperturbative processes in QCD.
 We have also demonstrated the important role of the BPT with confinement in establishing an internally consistent   theory of QCD without extra fitting parameters.
 We conclude  our exposition above with the assertion that the confinement mechanism, discovered in \cite{3,4,5,6} and strongly supported by lattice and experimental data, gives
 an important basis and stimulus for development of the internally consistent QCD both in the perturbative and nonperturbative aspects.
 .


The author is grateful to A.M.Badalian for very useful discussions and important suggestions and to A.L.Kataev
for a useful correspondence.

\vspace{2cm}

{\bf Appendix A1.}
{\bf The QCD vector coupling $\alpha_V(r)$}\\

 \setcounter{equation}{0} \def\theequation{A1.\arabic{equation}}

The spin-average masses of all mesons are described by the universal static potential $V(r)$, taken in the linear + gluon-exchange (GE) form,
\be
V(r)= \sigma r  - \frac{4}{3} \frac{\alpha(r)}{r}.
\label{A1.}
\ee
Here the vector coupling $\alpha_V(r)$ in the coordinate space is defined via the vector coupling in the momentum space
$\alpha_V(Q^2)$ as
\be
\alpha_V(r) = \frac{2}{\pi} \int^\infty_0 {\rm d}Q \frac{\sin(Qr)}{Q} \alpha_V(Q^2),
\label{A2.}
\ee
where the integration goes over three regions: small momenta $Q\la 1.5$~GeV, where $\alpha_V(Q^2)$ is defined by $\Lambda_V(n_f=3)$; second interval,  1.5~GeV~$\leq Q\leq m_b$ ($m_b\cong 4.20$~GeV is the $b-$ quark current mass), where the coupling is defined by $\Lambda_V(n_f=4)$, and the region of large momenta,  $q\geq m_b$, where $\Lambda_V(n_f=5)$ has to be used. This vector coupling in the coordinate spaces will be called the compound $\alpha_V^c(r)$. It can be compared with $\alpha_V(r,n_f=3)$, calculated via $\alpha_V(Q,n_f=3)$, defined by the QCD constant
$\Lambda_V(n_f=3)$  over the whole region of the integration. The calculations show that the compound coupling $\alpha_V^c(r)$ coincides with precision accuracy with $\alpha_V(r,n_f=3)$
at all distances with exception of very small $r\la 0.1$~fm, where the compound coupling is a bit smaller and this difference has to be taken into account only for the bottomonium ground state
$\Upsilon(1S)$. Here in calculations we use the IR regulator $M_B=1.15$~GeV.

The vector constants $\Lambda_V(n_f)$ are expressed via $\Lambda_{\overline{MS}}(n_f)$ as $\Lambda_V(n_f)=\Lambda_{\overline{MS}}(n_f) \exp(\frac{a_1}{2\beta_0})$ \cite{15} and hence
$\Lambda_V(n_f=3)=1.4753~\Lambda_{\overline{MS}}(n_f=3)$). The value $\Lambda_{\overline{MS}}(n_f=3)=332(19)$~MeV, is now well established in lattice QCD \cite{17}. It gives

\be
\Lambda_V(n_f=3) = 492(28)~{\rm MeV}.
\label{eq.A3.}
\ee
Very close value, $\Lambda_{\overline{MS}}(n_f=3)=339(10)$~MeV, was also found in the analysis of experimental data in \cite{16}.

Now we can define the characteristic scales $r_i,~(r_0,r_1,r_2)$ of the static potential (the forth) and compare them with the lattice results \cite{17,18}.
Firstly, with the use of (\ref{A2.})  we calculate the two-loop coupling $\alpha_V(r,n_f=3)$ and the constants $C_i$, defined  at three points $r_i, i=0,1,2$,
\be
 C_i = r_i^2 \frac{\partial V(r)}{\partial r}; ~~C_0=1.65,~ C_1=1.0, ~C_2=0.50.
\label{eq.A4.}
\ee
In the coupling  $\alpha_V(r,n_f=3)$ we use $M_B=1.15$~GeV and  obtains
$$
r_0=2.319~{\rm GeV}^{-1}=0.459~{\rm fm};~~$$$$ r_1=1.530~{\rm GeV}^{-1}= 0.303~{\rm fm};~~$$\be ~~r_2=0.70~{\rm GeV}^{-1}= 0.139~{\rm fm},
\label{eq.A5.}
\ee
for which  the following $\alpha_V(r_i)$ and the derivatives $\alpha_V'(r_i)$ are calculated,
$$
\alpha(r_2)=0.4479, ~\alpha_V'(r_2)=0.1967;$$$$ \alpha_V(r_1)= 0.49715, ~\alpha_V'(r_1) =0.1363;$$\be \alpha_V(r_0)=0.5954,~ \alpha_V'(r_o)= 0.0840.
\label{eq.A6.}\ee
Now we give also $\alpha(\lambda_0)$ at the important point $\lambda_0=1.0$~GeV$^{-1}$=0.20 fm,
\be
\alpha(0.20~\rm fm) = 0.497,
\label{eq.A7.}
\ee
which is close to $\alpha_V(r=0.20$~fm)=0.567, derived in (\ref{17}). Notice that the calculated $r_0$ and $r_1=0.303$~fm are in good agreement with the lattice result,
$r_1=0.3106$~fm in \cite{36}).
Also in our calculations the ratio
\be
\frac{r_1}{r_2}= 2.186,
\label{eq.A8.}
\ee
is in good agreement with the same ratio, defined in the lattice calculations, $\frac{r_1}{r_2}=2.198(9)$ \cite{36}.

\end{document}